\documentclass[preprint,showpacs,preprintnumbers,amsmath,amssymb]{revtex4}

\usepackage{graphicx}
\usepackage{epsfig}		
\usepackage{dcolumn}
\usepackage{bm}

\def\0{\mbox{\tiny $0$}}
\def\1{\mbox{\tiny $1$}}
\def\2{\mbox{\tiny $2$}}
\def\3{\mbox{\tiny $3$}}
\def\4{\mbox{\tiny $4$}}
\def\5{\mbox{\tiny $5$}}
\def\6{\mbox{\tiny $6$}}
\def\7{\mbox{\tiny $7$}}
\def\8{\mbox{\tiny $8$}}
\def\9{\mbox{\tiny $9$}}

\def\k{\mbox{\tiny $k$}}
\def\kk{\mbox{\small $k$}}

\def\foh{\mbox{\tiny $\frac{1}{2}$}}
\def\f14{\mbox{\tiny $\frac{1}{4}$}}

\def\l{\mbox{\tiny $l$}}

\def\s{\mbox{\tiny $s$}}
\def\sp{\mbox{\tiny $s^{\prime}$}}
\def\x{\mbox{\tiny $x$}}
\def\y{\mbox{\tiny $y$}}

\def\mi{\mbox{\tiny $-$}}
\def\ig{\mbox{\tiny $=$}}
\def\pl{\mbox{\tiny $+$}}

\def\bb#1{\mbox{\footnotesize $(#1)$}}
\begin{document}


\title{Chiral oscillations in terms of the {\em zitterbewegung} effect}

\author{A. E. Bernardini}
\affiliation{Instituto de F\'{\i}sica Gleb Wataghin, UNICAMP,\\
PO Box 6165, 13083-970, Campinas, SP, Brasil.}
\email{alexeb@ifi.unicamp.br}

\date{January, 2006}

\begin{abstract}
We seek the {\em immediate} description of chiral oscillations through the trembling motion obtained by the velocity (Dirac) operator $\vec{\alpha}$.
By taking into account the complete set of Dirac formalism solutions which results in a free propagating Dirac wave packet composed by positive and negative frequency components, we report about the well-established {\em zitterbewegung} results and indicate how chiral oscillations can be expressed through the well know quantum oscillating variables.    
We conclude with the interpretation of chiral oscillations as space coordinate very rapid oscillation projections onto the longitudinally decomposed direction of the motion.
\end{abstract}

\maketitle


After establishing the fundamental framework for relativistic quantum mechanics of ferminonic
particles \cite{Dir28,Dir30}, it became soon clear that the Dirac formalism, in spite of its notorious success in obtaining the energy levels of the hydrogen atoms, could exhibit a plenty of {\em supposed} inconsistencies, namely the presence of negative frequency solutions, the Klein paradox \cite{Kle29}
and the {\em zitterbewegung} ({\em ZWB}) phenomenon \cite{Sch30}.
In first quantization, besides the existence of
the spin one-half particle, under certain conditions, the relativistic Dirac formalism predicts an oscillating time dependence in the average of space coordinate variable.
This phenomenon, which is called as {\em ZWB}, was formerly noticed by Schroedinger \cite{Sch30} as a
consequence of the non commutative relation between the space coordinate operator $\vec{x}$ and the Hamiltonian $\mathcal{H}$.
The existence of the {\em ZWB} effect is no stranger than the existence of negative frequency solutions
given that such a trembling motion is indeed only manifest for wave functions with significant superposition between positive and negative frequency components of the Dirac fields.
For instance, in Hydrogen-like systems, the electron might be driven by violent quantum fluctuations in its intrinsic space coordinate as to become sensitive to an effective potential, effectively explained by the Darwin term of Coulomb potentials \cite{Zub80}\footnote{One could point out the possibility of suppressing {\em ZWB} oscillations by redefining
the coordinate variable via Foldy-Wouthuysen transformation.
However, it would lead to very complicated space coordinate variables.}.
Novel motivations for identifying the physical
observables which coexist with the superposition between positive and negative frequency Dirac spinor solutions have been considered in the literature \cite{Bra99,Rup00,Wan01,Bol04}.
By following a similar line of reasoning as that applied for computing the net effect of rapid oscillations of the space coordinate, it is reasonable to verify that the expectation value of the Dirac chiral operator $\Gamma^{\5}$ also exhibits such an oscillatory feature.
In particular, the formalism with Dirac wave packets \cite{Zub80,Ber04} also supports the framework of chiral oscillations \cite{DeL98} and, for flavor-changing interactions, neutrinos with positive chirality are decoupled from the neutrino absorbing charged weak currents \cite{DeL98}.
Therefore chiral coupled to flavor oscillations introduce modifications to the standard flavor conversion formula \cite{Ber05}.

Due to this tiny relation between {\em ZWB} and chiral oscillations, the question we shall answer in this manuscript is related to the computation of chiral oscillations in
terms of the {\em ZWB} motion: a chiral conversion mechanism is coupled with the {\em ZWB} motion in a way that they cannot exist independently.
That is, fast oscillations of the space coordinate variable can be decomposed into transversal and longitudinal polarization vectors, and chiral oscillations can be understood as space coordinate fast oscillation projections onto
the momentum direction - an important issue in the extended context of quantum oscillation phenomena.

Straightforward manipulations concern with a free propagating {\em bi-spinor} particle such that the covariant free particle Dirac formalism can be introduced through
\begin{equation}
\left(i \Gamma^\nu \partial_{\nu} - m \right)\varphi \bb{x} = 0,
\label{00}
\end{equation}
where $x = (t, \vec{x})$ one has considered the Dirac representation,
\cite{Sak87,Zub80,Wei95}m assuming that $c = \hbar = 1$.
Free particle solutions are given in terms of plane waves through
$\varphi\bb{x} = \varphi_{_{\pl}}\bb{x} + \varphi_{_{\mi}}\bb{x}$ with
$\varphi_{_{\pl}}\bb{x} = e^{[-  i \, p \, x]} \, u\bb{p}$ for positive frequencies,
and
$\varphi_{_{\mi}}\bb{x} = e^{[+ i \, p \, x]} \, v\bb{p}$ for negative frequencies
where $p$ is the relativistic {\em quadrimomentum}, $p = (E, \vec{p})$ with
$E^{\2} = m^{\2}+ \vec{p}^{\2}$,
and the free propagating mass-eigenstate spinors are introduced through \cite{Pes95}
{\small\begin{eqnarray}
u_{\s}\bb{p} &=& \frac{\Gamma^\nu p_\nu + m}{\left[2\,E\,(m+E)\right]^{\frac{1}{2}}}  \, u_{\s}\bb{m,\,0} = \left(\begin{array}{r} \left(\frac{E+m}{2E}\right)^{\foh} \eta_{\s}\\ \\ \frac{\vec{\sigma}.\vec{p}}{\left[2\,E\,(E+m)\right]^{\foh}} \eta_{\s} \end{array}\right),\nonumber\\
v_{\s}\bb{p} &=& \frac{-\Gamma^\nu p_\nu + m}{\left[2\,E\,(m+E)\right]^{\foh}} \, v_{\s}\bb{m,\,0} = \left(\begin{array}{r} \frac{\vec{\sigma}.\vec{p}}{\left[2\,E\,(E+m)\right]^{\foh}} \eta_{\s}\\ \\\left(\frac{E+m}{2E}\right)^{\foh} \eta_{\s} \end{array}\right),~~~
\label{02}
\end{eqnarray}}\normalsize
with $\overline{u}\bb{p}$ (or $\overline{v}\bb{p}$) set as
$\overline{u}\bb{p} = u^{\dagger}\bb{p}\Gamma^{\0}$ (or $\overline{v}\bb{p} = v^{\dagger}\bb{p}\Gamma^{\0}$).
The usual procedure \cite{Sak87} consists in assuming a Dirac wave packet solution of Eq.~(\ref{01}) as
\begin{eqnarray}
\varphi\bb{t, \vec{x}}
= \int\hspace{-0.1 cm} \frac{d^{\3}\hspace{-0.1cm}\vec{p}}{(2\pi)^{\3}}
\sum_{\s \ig \1,\2}\{b_{\s}\bb{p}u_{\s}\bb{p}\, \exp{[- i\,E\,t]}
+ d^*_{\s}\bb{\tilde{p}}v_{\s}\bb{\tilde{p}}\, \exp{[+i\,E\,t]}\}
\exp{[i \, \vec{p} \cdot \vec{x}]}~~
\label{03}
\end{eqnarray}
where $\overline{u}\bb{p}$ (or $\overline{v}\bb{p}$) is set as
$\overline{u}\bb{p} = u^{\dagger}\bb{x}\Gamma^{\0}$ (or $\overline{v}\bb{p} = v^{\dagger}\bb{p}\Gamma^{\0}$)
and $\tilde{p} = \bb{E, -\vec{p}}$.
Setting the boundary condition over $\varphi\bb{0,\mathbf{x}}$
through the Fourier transform of the weight function as
\begin{equation}
\varphi\bb{\vec{p}-\vec{p}_{\0}}\,w \,=\,
\sum_{\s \ig \1,\2}\{b_{\s}\bb{p}u_{\s}\bb{p} + d^*_{\s}\bb{\tilde{p}}v_{\s}\bb{\tilde{p}}\}
\label{002}
\end{equation}
one obtains
\begin{equation}
\varphi\bb{0, \vec{x}}
= \int\hspace{-0.1 cm} \frac{d^{\3}\hspace{-0.1cm}\vec{p}}{(2\pi)^{\3}}
\varphi\bb{\vec{p}-\vec{p}_{\0}}\exp{[i \, \vec{p} \cdot \vec{x}]}
\,w
\label{003}
\end{equation}
$w$ set as some constant normalized spinor.
The coefficients $b_{\s}\bb{p}$ and $d^*_{\s}\bb{\tilde{p}}$ are thus obtained by
using the orthogonality of Dirac bi-spinors.
For {\em any} initial state $\varphi\bb{0,\mathbf{x}}$ introduced through Eq.~(\ref{003}),
the coefficient of negative frequency solution 
$d^*_{\s}\bb{\tilde{p}}$ provides a non-vanishing contribution to the time-dependent wave packet.
The complete set of Dirac solutions is a necessary demand to build the wave packet with a semi-analytical profile.
Whether one considers a momentum distribution function built through a delta function limit, and with a constant spinor $w$ corresponding to a positive energy mass-eigenstate with momentum
$\vec{p}$, the contribution due to the negative frequency components 
$d^*_{\s}\bb{\tilde{p}}$ shall vanish.

With the Dirac Hamiltonian of the free propagating particle written as $\mathcal{H} = \vec{\alpha}\cdot \vec{p} + \beta m$, with $\vec{\alpha} = \sum_{\k \ig \1}^{\3} \alpha_{\k}\hat{\kk} = \sum_{\k \ig \1}^{\3} \Gamma_{\0}\Gamma_{\k}\hat{\kk}$
and $\beta = \Gamma_{\0}$, one easily verifies the conditions for a given observable $\mathcal{O}$ be a constant of the motion.
Through the Heisenberg formalism one has
\begin{equation} 
\frac{d~}{dt}\langle\mathcal{O}\rangle = i \langle\left[ \mathcal{H} ,\mathcal{O} \right]\rangle
+ \langle\frac{\partial \mathcal{O}}{\partial t}\rangle 
\label{09},
\end{equation}
 
For instance, the free propagating particle momentum can be identified as a conserved quantity since
\begin{equation} 
\frac{d~}{dt}\langle\vec{p}\rangle \,=\, i\langle\left[ \mathcal{H} , \vec{p}\right]\rangle\,=\,0 
\label{10},
\end{equation}
Likewise, the particle velocity comes out as
\begin{equation} 
\frac{d~}{dt}\langle\vec{x}\rangle \,=\, i\langle\left[ \mathcal{H} , \vec{x}\right]\rangle\,=\,\langle\vec{\alpha}\rangle 
\label{11}
\end{equation}
At first glance, it appears quite reasonable \cite{Sak87} to calculate only positive frequency component ($\varphi_{_{\pl}}\bb{x}$) averaged values as  
\begin{eqnarray} 
\langle\vec{\alpha}\rangle _{\pl}\bb{t} &=&
\int\hspace{-0.1 cm} d^{\3}\hspace{-0.1cm}\vec{x}\varphi_{\pl}^{\dagger}\bb{x} \,\vec{\alpha} \, \varphi_{\pl}\bb{x}\nonumber\\
&=&
\int\hspace{-0.1 cm} \frac{d^{\3}\hspace{-0.1cm}p}{(2\pi)^{\3}}\frac{\vec{p}}{E}\sum_{\s \ig \1,\2} |b_{\s}\bb{p}|^{\2}.
\label{12}
\end{eqnarray}
It gives the expectation value of the particle's velocity $\frac{\vec{p}}{E}$. 
In particular, one might notice that the eigenvalues of $\alpha_{\k}$ are $\pm 1$ and they corresponds to $\pm c$.
However, massive particles cannot exhibit, as an approached classical feature, velocities equal to $\pm c$.
Given that $\alpha_{\k}$ and $\alpha_{\l}$ do not commute one each other for $k \neq l$, the measurement of the $x$ projection of the particle's velocity is not consistent with the measurement of the corresponding $y$ projection - which is unusual, given that one knows that $p_{\x}$ and $p_{\y}$ commute.

The explanation for such an apparent misconception is given in therms of the plane wave solutions from Eq.~(\ref{02}) which  are not eigenfunctions of $\alpha_{k}$.
Therefore, the velocity operator, $\vec{\alpha}\bb{t}$ does not correspond to a constant of motion, namely,
\begin{equation} 
\frac{d~}{dt}\langle\vec{\alpha}\rangle \,=\, i\langle\left[ \mathcal{H} , \vec{\alpha}\bb{t}\right]\rangle
											\,=\, 2 \, i \left(\langle\vec{p}\rangle - \langle\vec{\alpha}\mathcal{H}\rangle\right)
\label{13}
\end{equation}
identified as a differential representation of $\vec{\alpha}\bb{t}$.
Reminding that $\vec{p}$ and $\mathcal{H}$ are
constants of the motion, one can solve Eq.~(\ref{13}) as to have \cite{Sak87}
\begin{equation} 
\langle\vec{\alpha}\rangle\bb{t} =
\langle \vec{p}\mathcal{H}^{\mi\1}\rangle 
+\langle\left(\vec{\alpha}\bb{0} -  \vec{p}\mathcal{H}^{\mi\1} \right)
e^{[-2\,i\, \mathcal{H} \, t]}\rangle~~~~
\label{14}
\end{equation}
 
Considering that the spin angular momentum operator related to the Dirac {\em bi-spinor} is given by
$\vec{\Sigma} = \Gamma^{\5} \vec{\alpha}$ is not a constant of the motion since 
\begin{equation} 
\frac{d~}{dt}\langle\vec{\Sigma}\rangle \,=\,i\langle\left[ \mathcal{H} , \vec{\Sigma} \right]\rangle
									  \,=\, -2 (\langle\vec{\alpha}\rangle \times \vec{p})
									  \label{15}
\end{equation}
the particle's {\em helicity} is introduced through the projection of the spin angular momentum onto the momentum vector as  $h = \frac{1}{2}\vec{\Sigma}\cdot\hat{\vec{p}}$, such that
\begin{equation} 
\frac{d~}{dt}\langle h \rangle \,=\, i\langle\left[ \mathcal{H} , h\right]\rangle\,=\,- \langle(\vec{\alpha} \times \vec{p})\cdot\hat{\vec{p}}\rangle \,=\ 0
\label{16}
\end{equation}
and the {\em helicity} is, therefore, a constant of motion.
Otherwise, the chiral operator $\Gamma^{\5}$, can be shown to be a time-dependent quantity \cite{DeL98}, 
\begin{equation} 
\frac{d~}{dt}\langle\Gamma^{\5}\rangle \,=\, i\langle\left[ \mathcal{H} , \Gamma^{\5}\right]\rangle\,=\,2\, i\, m \langle\Gamma^{\0}\Gamma^{\5}\rangle
\label{17},
\end{equation}
as helicity eigenstates can be read as chiral eigenstates uniquely in the ultra-relativistic limit (mass $m = 0$) \cite{Pes95}. 
The effective value of Eq.~(\ref{17}) appears only when both positive and negative components are taken into account to compose a Dirac wave packet, i. e. 
the non-vanishing averaged value of $\langle\Gamma_{\0}\Gamma_{\5}\rangle$ is revealed by the superposition involving the complete set of Dirac spinor solutions.

At time $t = 0$, the coefficients $b_{\s}\bb{p}$ and $d^*_{\s}\bb{p}$ used in the construction of the Dirac wave packet $\varphi\bb{x}$ can be chosen
to provide a negative (positive) chiralitly eigenstate \cite{Ber05,Ber05A} or, in the same way,
to provide a helicity eigenstate (when $h u_{\s}\bb{p}(v_{\s}\bb{p}) \equiv \pm \frac{1}{2}u_{\s}\bb{p}(v_{\s}\bb{p})$).
Once one has assumed that the initial chiral eigenstate\footnote{Neutrinos are supposedly produced as chiral eigenstates
via weak interactions.} $\varphi\bb{0, \vec{x}}$ is not only a superposition of momentum eigenstantes weighted by a momentum distribution
of Eq.~(\ref{002}) centered around $\vec{p}_{\0}$, but also a helicity (constant of the motion) eigenstate obtained through the production of a spin-polarized particle, 
which formally occurs when one assumes that the constant spinor $w$ in the wave packet expression (\ref{003})
is a simultaneous eigenspinor of $\Gamma^{\5}$ and $h$\footnote{When
one establishes that $\varphi\bb{0, \vec{x}}$ is a $h$ and/or a $\Gamma^{\5}$ eigenstate, it refers to the choice of the fixed spinor $w$ in Eq.~(\ref{003}).
As a peculiarity, {\em broken} of the Lorentz symmetry is not specifically related to the choice of $w$, but more generically 
to the choice of the momentum distributions ${b_{\s}\bb{p}}$ and $d^*_{\s}\bb{\tilde{p}}$ written in terms of $w$ and
$\varphi\bb{\vec{p}-\vec{p}_{\0}}$.
It is not the choice of $w$ as a $h$ eigenstate which ruins the Lorentz invariance of Dirac wave packets,
but the general choice of the momentum distribution for constructing them and the effective way that they
(Dirac wave packets) appear in some averaged value integrals,
i. e. given that one has established an analytical shape for the momentum distribution
(as they have done in done in \cite{DeL98} and \cite{Zub80}),
it is stated that it is valid only for one specific reference frame, and the discussion of Lorentz invariance aspects
becomes a little more complicated since it touches more fundamental definitions.},   
we can make use of the following decomposition,
\begin{eqnarray} 
\frac{d~}{dt}\langle \Gamma_{\5}\rangle\bb{t} &=&
\frac{d~}{dt}\langle\vec{\alpha} \cdot \hat{p}\left(\vec{\Sigma} \cdot \hat{p}\right)\rangle\bb{t}
=\left(\frac{d~}{dt}\langle\vec{\alpha} \cdot \hat{p}\rangle\bb{t}\right)\langle\vec{\Sigma} \cdot \hat{p}\rangle\bb{t}
+ \langle \vec{\alpha} \cdot \hat{p}\rangle\bb{t}\,\left(\frac{d~}{dt}\langle \vec{\Sigma} \cdot \hat{p}\rangle\bb{t}\right)~~~~
\label{18}
\end{eqnarray} 
from which, when one substitutes Eqs.~(\ref{10}) and (\ref{17}), a subtle relation between the chirality operator $\Gamma_{\5}\bb{t}$ and the
velocity operator $\vec{\alpha}\bb{t}$ appears as
\begin{equation} 
\frac{d~}{dt}\langle \Gamma_{\5}\rangle\bb{t} \,=\,
(2 h)\,\left(\frac{d~}{dt}\langle\vec{\alpha}\rangle\bb{t}\right) \cdot \hat{p}.
\label{19}
\end{equation}
 
The time evolution of $\Gamma_{\5}$ presents an oscillating character which can be interpreted as consequence of the oscillating features related to the trembling motion associated to $\vec{\alpha}\bb{t}$.
Eqs.~(\ref{13}) and (\ref{19}) lead to an evinced dependence of $\Gamma_{\5} \bb{t}$ on the
momentum, such that
$\vec{\alpha}\bb{t}$,
\begin{equation} 
\frac{d~}{dt}\langle\Gamma_{\5}\rangle \bb{t} =
 4 \, i \, h \,\left(\vec{p} - \langle\vec{\alpha}\bb{t}\,\mathcal{H}\rangle\right)\cdot \hat{p}.
\label{20}
\end{equation}
 
Given that $\langle h \rangle$, $\vec{p}$ and $\mathcal{H}$ are time independent, one concludes that there will not be chiral oscillations without the quivering motion of the space coordinate.
The constraint between $\Gamma_{\5}\bb{t}$ and $\vec{\alpha}\bb{t}$ operators become more interesting when the
complete expression for the current density $\overline{\varphi}\bb{x} \Gamma_{\nu}\, \varphi\bb{x}$ (which leads to the averaged value of
$\vec{\alpha}\bb{t}$) is considered.
Through the {\em Gordon decomposition} \cite{Pes95}, one then has
\begin{eqnarray} 
\overline{\varphi}\bb{x} \Gamma_{\nu} \varphi\bb{x} &=& -\frac{i}{2m}\left[\left(\partial^{\nu}\overline{\varphi}\bb{x}\right)
\varphi\bb{x} - \overline{\varphi}\bb{x} \left(\partial^{\nu}\varphi\bb{x}\right)\right]\nonumber\\
&&+ \frac{1}{2m}\partial^{\kappa}\left(\overline{\varphi}\bb{x} \sigma_{\nu\kappa} \varphi\bb{x}\right),
\label{22}
\end{eqnarray} 
where  $\sigma_{\nu\kappa} = \frac{i}{2}[\Gamma_{\nu},\Gamma_{\kappa}]$.
The integration of the vector components of Eq.~(\ref{22}) over position results into
\begin{eqnarray} 
\int\hspace{-0.1 cm} d^{\3}\hspace{-0.1cm}\vec{x}\,\varphi^{\dagger} \vec{\alpha}  \varphi &=&
\frac{1}{2m}\int\hspace{-0.1 cm} d^{\3}\hspace{-0.1cm}\vec{x} \left\{-i \left[\overline{\varphi} \left(\vec{\nabla} \varphi\right) - \left(\vec{\nabla} \overline{\varphi}\right)  \varphi\right]\right.\nonumber\\
&&\left.+ \,\left[\vec{\nabla}\times \left(\overline{\varphi} \vec{\Sigma} \varphi\right) - i \partial_{t}\left(\overline{\varphi} \vec{\alpha} \varphi\right)\right]\right\}
\label{23}
\end{eqnarray} 
where the $x$ dependence has been suppressed from the notation.
Using Eq.~(\ref{03}), the decomposed components of
$\langle\vec{\alpha}\rangle$ are set as
\begin{eqnarray} 
\int\hspace{-0.1 cm} d^{\3}\hspace{-0.1cm}\vec{x}\,\vec{\nabla}\times \left(\overline{\varphi} \vec{\Sigma}  \varphi\right) = 0
\label{24},
\end{eqnarray}
\begin{widetext}
\begin{eqnarray} 
\lefteqn{
\frac{i}{2m}\int\hspace{-0.1 cm} d^{\3}\hspace{-0.1cm}\vec{x}\,
\left[\overline{\varphi} \left(\vec{\nabla} \varphi\right) - \left(\vec{\nabla} \overline{\varphi}\right)  \varphi\right] =
\int\hspace{-0.1 cm} \frac{d^{\3}\hspace{-0.1cm}p}{(2\pi)^{\3}}\left\{\frac{\vec{p}}{E}
\sum_{\s \ig \1,\2} \left[|b_{\s}\bb{p}|^{\2}+|d_{\s}\bb{p}|^{\2}\right]
\right.}\nonumber\\
&& \left. ~~~~~~~ 
+\sum_{\s \ig \1,\2}\left(\frac{m}{E}-\frac{E}{m}\right)a_{\s}\hat{p} \left[b^*_{\s}\bb{p}\,d^*_{\s}\bb{\tilde{p}}\, e^{[+2 \,i\,E\,t]} \,-\, d_{\s}\bb{p}\,b_{\s}\bb{\tilde{p}}\, e^{[-2 \,i\,E\,t]}\right]\right\}
\label{25},
\end{eqnarray} 
with $\tilde{p} = (E,-{\vec{p}})$, and assuming that $a_{\s} = \eta_{\sp}\, \vec{\sigma} \cdot \hat{p} \,\eta_{\s} = (-1)^{\s \pl \1} \delta_{s s^{\prime}}$, and
\begin{eqnarray} 
-\frac{i}{2m}\int\hspace{-0.1 cm} d^{\3}\hspace{-0.1cm}\vec{x}\,\partial_{t}\left(\overline{\varphi} \vec{\alpha} \varphi\right) &=& 
\int\hspace{-0.1 cm} \frac{d^{\3}\hspace{-0.1cm}p}{(2\pi)^{\3}}\left\{\sum_{\s \ig \1,\2}\left(\frac{E}{m}\right)a_{\s}\hat{p} \left[b^*_{\s}\bb{p}\,d^*_{\s}\bb{\tilde{p}}\, e^{[+2 \,i\,E\,t]} \,-\, d_{\s}\bb{p}\,b_{\s}\bb{\tilde{p}}\, e^{[-2 \,i\,E\,t]}\right]\right.\nonumber\\
&&~~~~~~~~~~~ + \left.\sum_{\s \neq \sp} \hat{n}_{\s}  \left[b^*_{\s}\bb{p}\,d^*_{\sp}\bb{\tilde{p}}\, e^{[+2 \,i\,E\,t]} \,-\, d_{\s}\bb{p}\,b_{\sp}\bb{\tilde{p}}\, e^{[-2 \,i\,E\,t]}\right]\right\}
\label{26}
\end{eqnarray} 
\end{widetext}
where $\hat{n}_{\1(\2)} = \hat{1}\pm i \hat{2}$ when $\hat{p} = \hat{3}$ and
the unitary vectors $\hat{1}$, $\hat{2}$ and $\hat{3}$ describe mutually orthogonal components.

Eq.~(\ref{24}) allows one to verify that the {\em ZWB} does not get a contribution
from the intrinsic spin dependent ($\vec{\Sigma}$) magnetic moment component which couples with
external magnetic fields $\vec{B}\bb{x}$.
In fact, the {\em ZWB} originates from the current strictly related to the internal electric moment.
To clear up this point, it is convenient to consider the modern and more precise interpretation of the {\em ZWB} \cite{Bar81,Riv02}.
From such a theoretical perspective, the {\em ZWB} for a Dirac particle is clearly related to the separation between the center of mass, which is related to the Foldy-Wouthuysen space coordinate operator, and the center of charge, which corresponds to Dirac's space coordinate operator $\vec{x}$.
It is the particle's charge at space coordinate $\vec{x}$ that is moving at the speed of light in circles of radius $\hbar/2 m c$ around the center of mass, such that the average value of this velocity is related to the linear momentum of the particle.
It is this separation, and thus the existence of an electric dipole moment with respect to the center of mass, which justifies the above statement.
For both particles and antiparticles, then in pure positive or negative energy states, this internal motion of the charge around the center of mass exists but there are no chiral oscillations, which is no longer obvious.
However, the production of a bi-spinor particle as a chiral eigenstate (for instance, a neutrino), imposes the wave packet space-time evolution obtained by the superposition of positive and negative frequency components of the Dirac formalism \cite{Ber04}.
It recovers the possibility of chiral oscillations for massive bi-spinor particles.
In this context, the coupling between chiral oscillations and the {\em zitterbwegung} is recovered.

By taking into account Eqs.~(\ref{25}-\ref{26}),
one can turn back to Eq.~(\ref{20}), which carries the main idea of this manuscript,
and observe that chiral oscillations can be essentially constructed in terms of the longitudinal components of
$\langle\vec{\alpha}\rangle$. By calculating the mean value of $\langle\vec{\alpha} \mathcal{H}\rangle$
and projecting it onto the momentum direction $\hat{p}$, one obtains
\begin{widetext}
\begin{eqnarray} 
\langle\vec{\alpha}\bb{t}\,\mathcal{H}\rangle\cdot \hat{p} &=
&\int\hspace{-0.1 cm} \frac{d^{\3}\hspace{-0.1cm}p}{(2\pi)^{\3}}\left\{\frac{\vec{p}}{E}\cdot \hat{p}
\sum_{\s \ig \1,\2} \left[(E)|b_{\s}\bb{p}|^{\2}+(E)|d_{\s}\bb{p}|^{\2}\right]
\right. \nonumber\\ 
&& \left. ~~~~~~~~~~~
+ \sum_{\s \ig \1,\2} \frac{m}{E}\,a_{\s} \left[(E)b^*_{\s}\bb{p}\,d^*_{\s}\bb{\tilde{p}}\, e^{[+ 2 \,i\,E\,t]} \,-\,(E) d_{\s}\bb{p}\,b_{\s}\bb{\tilde{p}}\, e^{[-2 \,i\,E\,t]}\right]\right\}\nonumber\\
&=& |\vec{p}| - 
\int\hspace{-0.1 cm} \frac{d^{\3}\hspace{-0.1cm}p}{(2\pi)^{\3}}\, m \,\sum_{\s \ig \1,\2}
a_{\s} \left[b^*_{\s}\bb{p}\,d^*_{\s}\bb{\tilde{p}}\, e^{[+2 \,i\,E\,t]} \,-\,d_{\s}\bb{p}\,b_{\s}\bb{\tilde{p}}\, e^{[-2 \,i\,E\,t]}\right]
\label{27},
\end{eqnarray} 
which can be substituted into (\ref{20}) as to give
\begin{eqnarray} 
\frac{d~}{dt}\langle\Gamma_{\5}\rangle \bb{t} &=&
\int\hspace{-0.1 cm} \frac{d^{\3}\hspace{-0.1cm}p}{(2\pi)^{\3}}\,
\frac{m}{E} \sum_{\s \ig \1,\2} (2 \,h \,a_{\s})  (2 \, i \, E) \left[b^*_{\s}\bb{p}\,d^*_{\s}\bb{\tilde{p}}\, e^{[+2 \,i\,E\,t]} \,-\, d_{\s}\bb{p}\,b_{\s}\bb{\tilde{p}}\, e^{[-2 \,i\,E\,t]}\right]
\label{28}
\end{eqnarray} 
in manner that the time dependence of the chiral operator could be written as
\begin{eqnarray} 
\langle\Gamma_{\5}\rangle \bb{t} &=&
\langle\Gamma_{\5}\rangle \bb{0} +
\int\hspace{-0.1 cm} \frac{d^{\3}\hspace{-0.1cm}p}{(2\pi)^{\3}}\,
\frac{m}{E}\,\sum_{\s \ig \1,\2}( 2\, h \, a_{\s})\, 
\left[d_{\s}\bb{p}\,b_{\s}\bb{\tilde{p}}\, \left(e^{[2 \,i\,E\,t]} - 1\right) + h.c.
\right]
\nonumber\\
&=&
\langle\Gamma_{\5}\rangle \bb{0} +
\int\hspace{-0.1 cm} \frac{d^{\3}\hspace{-0.1cm}p}{(2\pi)^{\3}}\,
\frac{m}{E}\,\sum_{\s \ig \1,\2} 
\left[d_{\s}\bb{p}\,b_{\s}\bb{\tilde{p}}\, \left(e^{[2 \,i\,E\,t]} - 1\right) + h.c.
\right]
\label{29}.
\end{eqnarray} 
\end{widetext}
since $2 h a_{\s}$ is unitary for positive and negative definite helicity states. 

The physical relevance of the above obtained results is discussed in \cite{DeL98},
in particular, for a gaussian state describing and initial wave function which has null averaged chirality: chiral {\em left-right} oscillations mutually cancel and there is no overall oscillation. 
It could be the origin of an apparent paradox.
Just to avoid to reproduce the same ideas already presented in \cite{DeL98}, while the cross sections are Lorentz invariant, the chiral probabilities are not, which suggests that probabilities measurements are chiral independent, ruining the physical significance of chiral oscillations.
The objection based upon the Lorentz invariance replied by the argument that, at any given
Lorentz frame, chiral oscillations are important because of the chiral projection form (V-A) of the charged weak currents.
The chiral probability variations produced by Lorentz transformations (even if $\Gamma^5$ commutes with the Lorentz generators) are reversely compensated by the wave function normalization and the Lorentz transformations of the weak interaction vector bosons and other interacting particles.

The above analysis also suggests that the preliminary tools for obtaining an expression for the neutrino spin-flipping in magnetic field can be related to chiral oscillations in the limit of a massless particle (ultra-relativistic limit).
By correctly differing the concepts of helicity and chirality, one can identify the origin and the influence of chiral oscillations and spin-flipping
in the complete flavor conversion formula.
In some previous manuscripts \cite{Ber04,Ber05,Ber05A}
we have also confirmed that the {\em bi-spinor} character of the particles changes the standard flavor oscillation
profiles by the presence of corrections due to a very high oscillation frequency dependence on the sum of energies which, in case of Dirac wave-packets, correspond to fine-tuning 
modifications that, however, are not effective in the UR regimes of propagating neutrinos in vacuum\cite{Kim93,MSW}.
The physical consequences in environments such as supernova can be theoretically studied \cite{Oli99}.
And also, it was observed that, in matter environments, neutrinos achieve an effective electromagnetic vertex which affects the flavor conversion process in a framework where conserving chirality can be obtained \cite{Oli96}.

Just as a remark about this connection with neutrino physics, it is also to be noted that in this kind of analysis one has to assume that neutrinos are
Dirac particles, thus making the positive-chiral component sterile. 
If neutrinos are Majorana particles \cite{Kim93}, they cannot have a magnetic moment, obviating the spin-flipping via magnetic field interactions
but still allowing the (vacuum) chiral conversion possibility via very rapid oscillations ({\em ZWB}).

To end up, given that neutrino electroweak interactions at the source and detector are presumably ({\em left}) chiral
$\left(\overline{\varphi} \Gamma^{\nu}(1 - \Gamma^{\5})\varphi W_{\nu}\right)$,
only the component with negative chirality contributes to the propagation.
Hence {\em ZWB} -like oscillations can take place in the scenario of neutrino oscillations.
It is remarkable that, in the standard treatment of neutrino flavor conversion mechanism, mass-eigenstate wave packets built with only positive frequency solutions is implicitly
assumed. Even if the standard oscillation predicts fiducial results when suitably interpreted, a more satisfactory description
involving bi-spinor structures require the use of the Dirac formalism for the propagating mass-eigenstates.
Consequently, the spinorial form and the superposition between positive and negative frequencies of the mass-eigenstate wave packets leads to the possibility
of chiral coupled with flavor oscillations \cite{Ber04}, even when it concerns with non-relativistic neutrinos.



\begin{thebibliography}{99}
\bibitem{Dir28}
P. A. M. Dirac, Proc. R. Soc. {\bf A117} 610 (1928).
\bibitem{Dir30}
P. A. M. Dirac, Proc. R. Soc. {\bf A126} 360 (1930).
\bibitem{Kle29}
O. Klein, Z. Physik {\bf 53} 157 (1929).
\bibitem{Sch30}
E. Schroedinger, Sitzber. Preuss. Akad. Wiss. Physik-Mat. {\bf 28} 418 (1930).
\bibitem{Zub80}
C. Itzykson and J. B. Zuber, {\em Quantum Field Theory}, (Mc Graw-Hill Inc., New York, 1980).
\bibitem{Bra99}
J. W. Braun, Q. Su and R. Grobe, Phys. Rev. {\bf A59}, 604 (1999).
\bibitem{Rup00}
S. Rupp, T. Sigg and M. Sorg, Int. J. Theor. Phys. {\bf 39}, 1543 (2000).
\bibitem{Wan01}
B. Thaller,
quant-ph/0409079
\bibitem{Bol04}
J. Bolte and R. Glaser, J. Phys. A: Math. Gen. {\bf 37}, 6359 (2004).
\bibitem{Ber04}
A. E. Bernardini and S. De Leo, Eur. Phys. J.{\bf C37}, 471 (2004).
\bibitem{DeL98}
S. De Leo and P. Rotelli, Int. J. Theor. Phys. {\bf 37}, 2193 (1998).
\bibitem{Ber05}
A. E. Bernardini and S. De Leo, Mod. Phys. Lett. {\bf A20}, 681 (2005)
\bibitem{Sak87}
J. J. Sakurai, {\em Advanced Quantum Mechanics}, (Addison-Wesley Publishing Company, New York, 1987).
\bibitem{Wei95}
S. Weinberg, {\em The Quantum Theory of Fields}, (Cambridge University Press, New York, 1995).
\bibitem{Ber05A}
A. E. Bernardini and S. De Leo, Phys. Rev. {\bf D71}, 076008-1 (2005)
\bibitem{Pes95}
M. E. Peskin and D. V. Schroeder, {\em An Introduction to Quantum Field Theory},
(Addison - Wesley Publishing Company, New York, 1995).
\bibitem{Bar81}
A. O. Barut and A. J. Bracken, {\em Phys. Rev.} {\bf D23}, 2454 (1981).
\bibitem{Riv02}
M. Rivas, {\em Kinematical theory of spinning particles}, Fundamental Theories of Physics - Vol. 116,  (Springer, Berlin, 2002).
\bibitem{Kim93}
C. W. Kim and A. Pevsner, {\em Neutrinos in Physics and Astrophysics}, (Harwood Academic Publishers, Chur, 1993).
\bibitem{MSW}
L. Wolfenstein, Phys. Rev. {\bf D17}, 2369 (1978); {\em ibid} {\bf D20}, 2634 (1979),\\
S. P. Mikheyev and A. Yu. Smirnov, Sov. J. Nucl. Phys. {\bf 42}, 913 (1986); Nuovo Cimento {\bf C9}, 17 (1986).
\bibitem{Oli99}
A. Ayala, J. C. D'Olivo and M. Torres, {\em Phys. Rev.} {\bf D59 RC}, 111901 (1999).
\bibitem{Oli96}
J. C. D'Olivo and J. F. Nieves, Phys. Lett {\bf B383}, 87 (1996).
\end{thebibliography}
\end{document}